\documentclass[aps,pra,showpacs,twocolumn,final,tightenlines]{revtex4}
\usepackage{amsmath}
\usepackage{color}
\usepackage{dcolumn}
\usepackage{bm}
\usepackage{amsfonts}
\usepackage{amssymb}
\usepackage{graphicx}

\begin{document}

\title{Collapse and revival dynamics of number-squeezed superfluids of
ultracold atoms in optical lattices}
\author{E. Tiesinga}
\affiliation{Joint Quantum Institute, National Institute of Standards and Technology and
University of Maryland, 100 Bureau Drive, Stop 8423 Gaithersburg, Maryland
20899-8423, USA}
\author{P. R. Johnson}
\affiliation{Department of Physics, American University, Washington DC 20016, USA}

\begin{abstract}
Recent experiments have shown a remarkable number of collapse-and-revival
oscillations of the matter-wave coherence of ultracold atoms in optical
lattices [Will \textit{et al.}, Nature 465, 197 (2010)]. Using a mean-field
approximation to the Bose-Hubbard model, we show that the visibility of
collapse-and-revival interference patterns reveal number squeezing of the
initial superfluid state. To describe the dynamics, we use an effective
Hamiltonian that incorporates the intrinsic two-body and induced three-body
interactions, and we analyze in detail the resulting complex pattern of
collapse-and-revival frequencies generated by virtual transitions to higher
bands, as a function of lattice parameters and mean-atom number. Our work
shows that a combined analysis of both the multiband, non-stationary
dynamics in the final deep lattice, and the number-squeezing of the initial
superfluid state, explains important characteristics of optical lattice
collapse-and-revival physics. Finally, by treating the two- and three-body interaction strengths, and the coefficients describing the initial superposition of number states, as free parameters in a fit to the experimental data it should be possible to go beyond some of the limitations of our model and obtain insight into the breakdown of the mean-field theory for the initial state or the role of nonperturbative effects in the final state dynamics.

\end{abstract}

\maketitle

\section{Introduction}

Optical lattices have become a powerful tool for studying the quantum
many-body states of ultracold atoms \cite{Morsch2006,Bloch2008}. In 1998
Jaksch \textit{et al.} \cite{Jaksch1998} suggested using ultracold atoms in
an optical lattice to simulate the Bose-Hubbard model \cite{Fisher1989}, and
in 2002 Greiner \textit{et al.} \cite{Greiner2002a} demonstrated Mott
insulator quantum phase transitions by imaging atoms released from a lattice
after a period of free expansion, which reveals the atomic momentum
distribution at the moment of release. Recently, the \textit{in-situ}
measurement of the position of atoms has become possible \cite%
{Nelson2007,Gemelke2009,Bakr2010}, and techniques to measure higher-order
correlation functions, thermodynamic properties such as pressure,
temperature, and transport properties have been developed \cite%
{Jordens2008,Hung2010,Ho2010}. In the Mott state pairs of atoms have been
studied spectroscopically \cite{Moritz2005,Ospelkaus2006,Widera2006}.

In parallel, techniques have been developed to create and exploit the
non-stationary dynamics of many-body states and their interactions. In
collapse and revival experiments \cite{Greiner2002b,Sebby2007,Will2010}, a
sudden increase in lattice depth projects an initial ground state into a
non-equilibrium state which is held in the deep lattice for a variable
period of time $t$. Coherence between confined atoms at the moment the
lattice is switched off results in an interference pattern in the
atom-density after a period of free expansion of the atom cloud \cite%
{Toth2008}. The interaction-induced non-stationary dynamics in the lattice
cause the degree of coherence to oscillate in time, which in turn makes the
visibility of imaged interference patterns, defined here as the ratio of
atoms in interference peaks divided by the total number of atoms, to
oscillate as a function of $t$.

In Ref.~\cite{Johnson2009} we showed that the crucial role of multiband
physics in collapse and revival dynamics can be elegantly described by
extending the Bose-Hubbard model to include effective three- and higher-body
interactions. After developing a new technique greatly reducing dephasing
from inhomogeneities, Will \textit{et al.}~\cite{Will2010} recently observed
tens of collapse-and-revival oscillations, which confirmed the predicted
higher-body effects \cite{Johnson2009}. The experimental data had
characteristics, however, not explained by higher-body interactions alone,
and Will \emph{et al.}~\cite{Will2010} suggested that the data reveals
additional information about the initial many-body state.

In this paper, we show that a combined analysis of both the number-squeezing
of the initial superfluid state and the non-stationary multiband dynamics in
the final deep lattice explains important characteristics of
collapse-and-revival physics. First, we show how the visibility of the
interference patterns depends on the degree of number-squeezing of the
initial superfluid in the shallow lattice, providing a new tool for
analyzing many-body states in optical lattices. Next, to describe the
dynamics in the deep lattice, we use an effective Hamiltonian that
incorporates the intrinsic two-body and induced three-body interactions, and
we analyze in detail the resulting complex pattern of collapse-and-revival
frequencies generated by virtual transitions to higher bands. In particular,
we show that the frequencies appearing the time evolution of the
interference pattern visibilities are robust under changes in mean atom
number, with number squeezing only affecting the spectral weights. Finally, by treating the two- and three-body interaction strengths, and the coefficients describing the initial superposition of number states, as free parameters in a fit to the experimental data it should be possible to go beyond some of the limitations of our model and obtain insight into the breakdown of the mean-field theory for the initial state or the role of nonperturbative effects in the final state dynamic

\section{Shallow lattice and initial state}

The initial state of atoms in an optical lattice is well described by the
ground state of the three-dimensional (3D) Bose-Hubbard model with its
parameters in a regime where tunneling cannot be neglected. The many-body
ground state to good approximation factorizes into a product of single-site
states \cite{Jaksch1998}. The breakdown of the factorization for small
lattices has been studied in \cite{Schachenmayer2011}.

In previous analyses of collapse-and-revival experiments~\cite%
{Greiner2002b,Sebby2007,Will2010}, the single-site state was modeled as a
coherent atom-number state $|\beta \rangle =e^{\beta b^{\dagger
}-\beta^{\ast }b}|0\rangle $ with mean atom number $\left\langle
n\right\rangle = \langle b^\dagger b\rangle=|\beta |^{2}$, where the
operators $b$ and $b^{\dagger }$ annihilate and create atoms in a lattice
site, respectively, and $|0\rangle $ is the zero atom vacuum. This model
captures basic aspects of the physics, but predicts contrasts and spectral
weights of the interference visibilities that differ significantly from the
data (e.g. compare the theoretical predictions in ~\cite{Johnson2009} to the
data in ~\cite{Will2010}). For example, the data in ~\cite{Will2010} show
minimal visibilities that are no smaller than $40\%$ for some system
parameters, before dephasing from inhomogeneities is significant, which is
much larger than the minimum visibility $e^{-4\left\langle n\right\rangle }$
predicted for collapse of a coherent state.

Below, we show that these deviations are associated with the number
squeezing predicted by a mean-field decoupling approximation \cite%
{Sheshadri1993,vanOosten2001}. This approximation gives the Hamiltonian for
each lattice site 
\begin{equation}
H_{i}=-zJ_{i}\phi (b+b^{\dagger })+zJ_{i}\phi ^{2}+\frac{1}{2}%
U_{2,i}\,b^{\dagger }b^{\dagger }bb-\mu \,b^{\dagger }b\,,
\end{equation}%
where $z=6$ is the number of nearest-neighbor sites in the 3D lattice, $%
J_{i} $ is the tunneling parameter, $U_{2,i}$ is the on-site atom-atom
interaction strength, and $\mu $ is the chemical potential, which sets the
mean atom number. The ground state is the lowest eigenstate of the
non-number conserving $H_{i}$ corresponding to a real-valued $\phi $ that
minimizes the energy. For this value of $\phi $ the superfluid order
parameter $\langle b\rangle =\phi $. A zero value for $\phi $ implies a Mott
state with definite atom number per lattice site. Reference \cite%
{Freericks1994} describes the breakdown of this approximation for 1D and 2D
systems.

Explicit expressions for $J_{i}$ and $U_{2,i}$ (see \cite{vanOosten2001})
contain overlap integrals of Wannier functions with the single-atom
Hamiltonian and the atom-atom interaction potential, respectively. We
consider lattices where the Wannier functions can be approximated by
harmonic oscillator solutions, and assume a regularized 3D delta function
potential for the two-body interactions between atoms \cite{Huang1957} with
a strength proportional to the scattering length of atoms colliding with
zero relative kinetic energy \cite{Chin2010}. These approximations allow us
to obtain useful analytic results, although fully quantitative modeling
should include the important role of the lattice anharmonicity.

\begin{figure}[tbp]
\includegraphics[width=8.5cm]{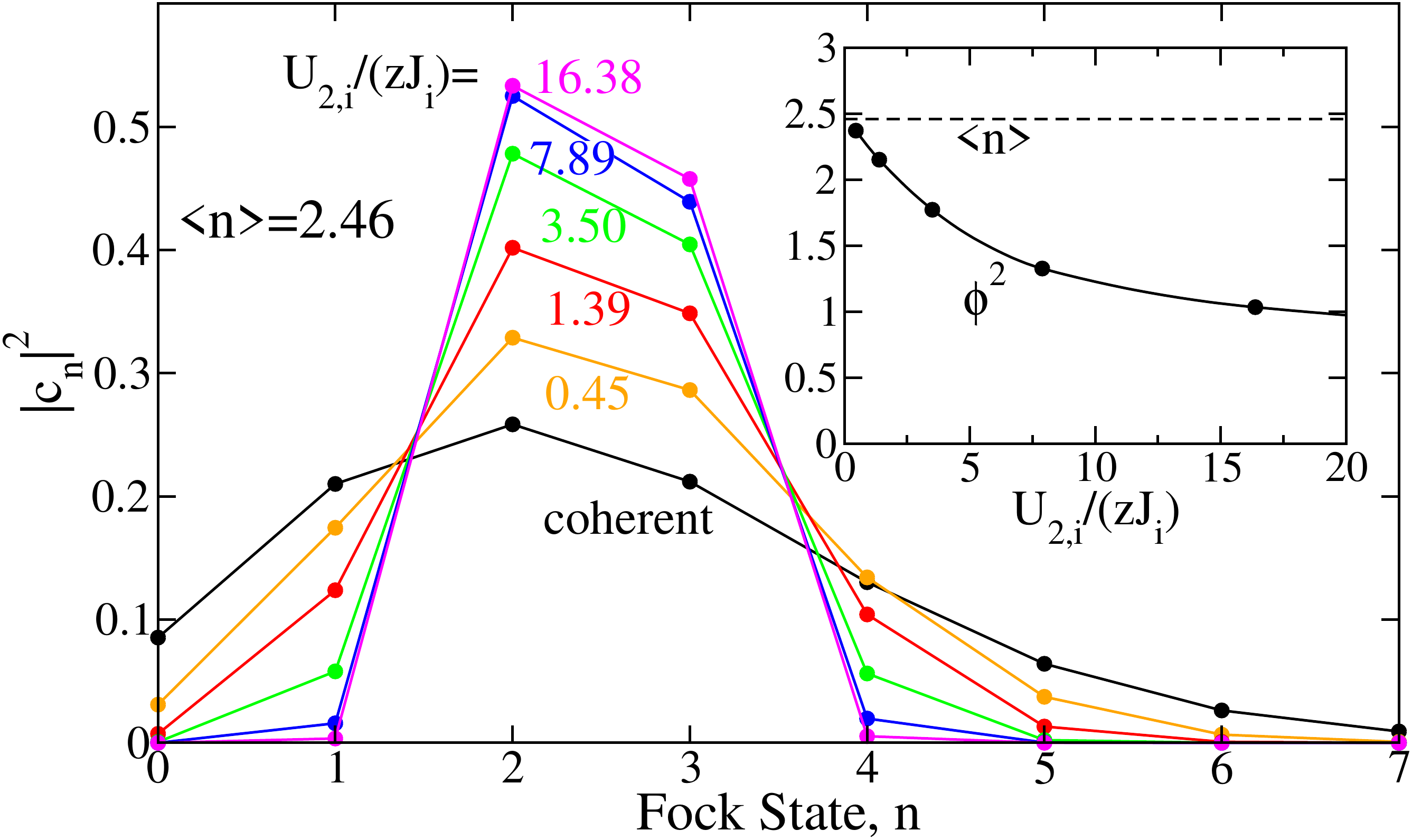}
\caption{(Color online) The probabilities $|c_{n}|^{2}$ for atom-number
states $|n\rangle $ at each lattice site calculated from a mean-field
decoupling approximation to the Bose-Hubbard model for several $\protect\xi %
=U_{2,i}/(zJ_{i})$ and fixed $\langle n\rangle =2.46$. The curve labeled
\textquotedblleft coherent\textquotedblright\ corresponds to a coherent
state with mean atom number $\langle n\rangle $. The inset shows the square
of the order parameter as a function of $\protect\xi $. The values of $%
\protect\xi $ corresponds to a linear increase of the lattice depth from 3$%
E_{R}$ to 11$E_{R}$ assuming $^{87}$Rb atoms and laser wavelength of 738 nm.
Here, $E_{R}$ is the single-photon recoil energy of a $^{87}$Rb atom.}
\label{initial}
\end{figure}

To obtain the ground state for different initial lattice depths we
numerically solve for the ground state of $H_{i}$ using the variational
solution $|\Psi _{i}\rangle =\sum_{n}c_{n}|n\rangle ,$ with Fock states $%
|n\rangle $ containing $n=0,1,2,\cdots $ atoms and amplitudes $c_{n}$. We
included $n$ up to 20. Figure \ref{initial} shows examples of probabilities $%
|c_{n}|^{2}$ for Fock state components $\left\vert n\right\rangle $ of the
ground state for several ratios of $\xi =U_{2,i}/(zJ_{i})$, but fixed $%
\langle n\rangle$. The probabilities for a coherent state $\left\vert \beta
\right\rangle $ with the same mean atom number are also shown. The ground
state is number squeezed with increasing $\xi $, and for large $\xi $ the
Fock states closest to $\langle n\rangle $ dominate. In the opposite limit $%
\xi \rightarrow 0$ it approaches a coherent state, however, number-squeezing
is readily apparent even for a shallow lattice as long as $\xi $ is large
enough that a single well-defined band exists. The inset shows the
superfluid order parameter versus $\xi $. For small $\xi $ the parameter $%
\phi $ approaches $\sqrt{\langle n\rangle }$, while for large $\xi $ the
order parameter slowly approaches zero. We next analyze how the frequencies
and visibilities of the collapse and revival oscillations depend on the
amplitudes $c_{n}$.

\section{Effective deep lattice dynamics}

Collapse and revival dynamics start after a sudden increase in lattice
depth, reducing the tunneling parameter to near zero. This operation is fast
enough that interactions can be neglected but slow enough to prevent
vibrational excitation in a lattice site. The state immediately after the
lattice transformation retains the form $\sum_{n}c_{n}|n\rangle $, although
the single particle basis functions change form slightly. Even with minimal
real single-particle excitation, however, multiband effects are critical due
to collision-induced virtual excitations to vibrationally-excited levels. We
have shown that these effects can be concisely included via the effective
Hamiltonian \cite{Johnson2009}%
\begin{equation}
H_{f}=\frac{1}{2}U_{2,f}a^{\dagger }a^{\dagger }aa+\frac{1}{6}%
U_{3,f}a^{\dagger }a^{\dagger }a^{\dagger }aaa+\mathcal{O}\left(
U_{2,f}^{3}\right) \,,
\end{equation}%
where $a^{\dagger }$($a$) creates (annihilates) an atom in a single
renormalized spatial mode in a site of the final optical lattice and $%
U_{2,f} $ and $U_{3,f}$ give the strength of effective two- and three-body
interactions.

The effective Hamiltonian $H_{f}$ applies to deep lattices with no
tunneling, such that the dynamics in each site is independent. This is valid
for the experiments in~\cite{Greiner2002b,Sebby2007,Will2010}, where the
tunneling and three-body recombination rates are 1-2 orders of magnitude
smaller than the three-body interaction strength. Recently, Ref.~\cite%
{Wolf2010} showed ways to measure tunneling rates from collapse and revival
experiments in less deep lattices. Effective three-body interactions have
also been shown to govern thermalization of a one-dimensional Bose gas \cite%
{Mazets2010}, and are important in Efimov physics \cite{Braaten2007}.

Renormalization fixes the value of $U_{2,f}$ to the measured energy of two
atoms held in a lattice site after the zero-point energy of two
non-interacting atoms is subtracted (Alternately, Ref.~\cite{Busch1998}
gives the exact analytical relationship between $U_{2,f}$ and the coupling
constant of the regularized 3D delta function potential for two harmonically
trapped atoms.) After renormalization, the effective three- and higher-body
interactions are predicted in terms of $U_{2,f}$. Ignoring anharmonicities
and assuming a harmonic potential with frequency $\omega _{f}$, we showed in
Ref.~\cite{Johnson2009} that to second-order $U_{3,f}$ is attractive and
given by 
\begin{equation}
U_{3,f}=-c_{f}\,U_{2,f}^{2}/(\hbar \omega _{f}),
\end{equation}%
with $c_{f}=4\sqrt{3}+6\ln (4/(2+\sqrt{3}))-6=1.344...$. The corrections due
to anharmonicities in the deep lattices considered here are less than 5\%.

The state evolves under $H_{f}$ for hold time $t$. Since $H_{f}$ is number
conserving, the state becomes $|\Psi _{f}(t)\rangle
=\sum_{n}c_{n}e^{-iE_{n}t/\hbar }|n\rangle $, where 
\begin{equation}
E_{n}=U_{2,f}n(n-1)/2+U_{3,f}n(n-1)(n-2)/6
\end{equation}
and $h=2\pi \hbar $ is Planck's constant.

\begin{figure}[t]
\vspace*{2mm} \includegraphics[width=8.5cm]{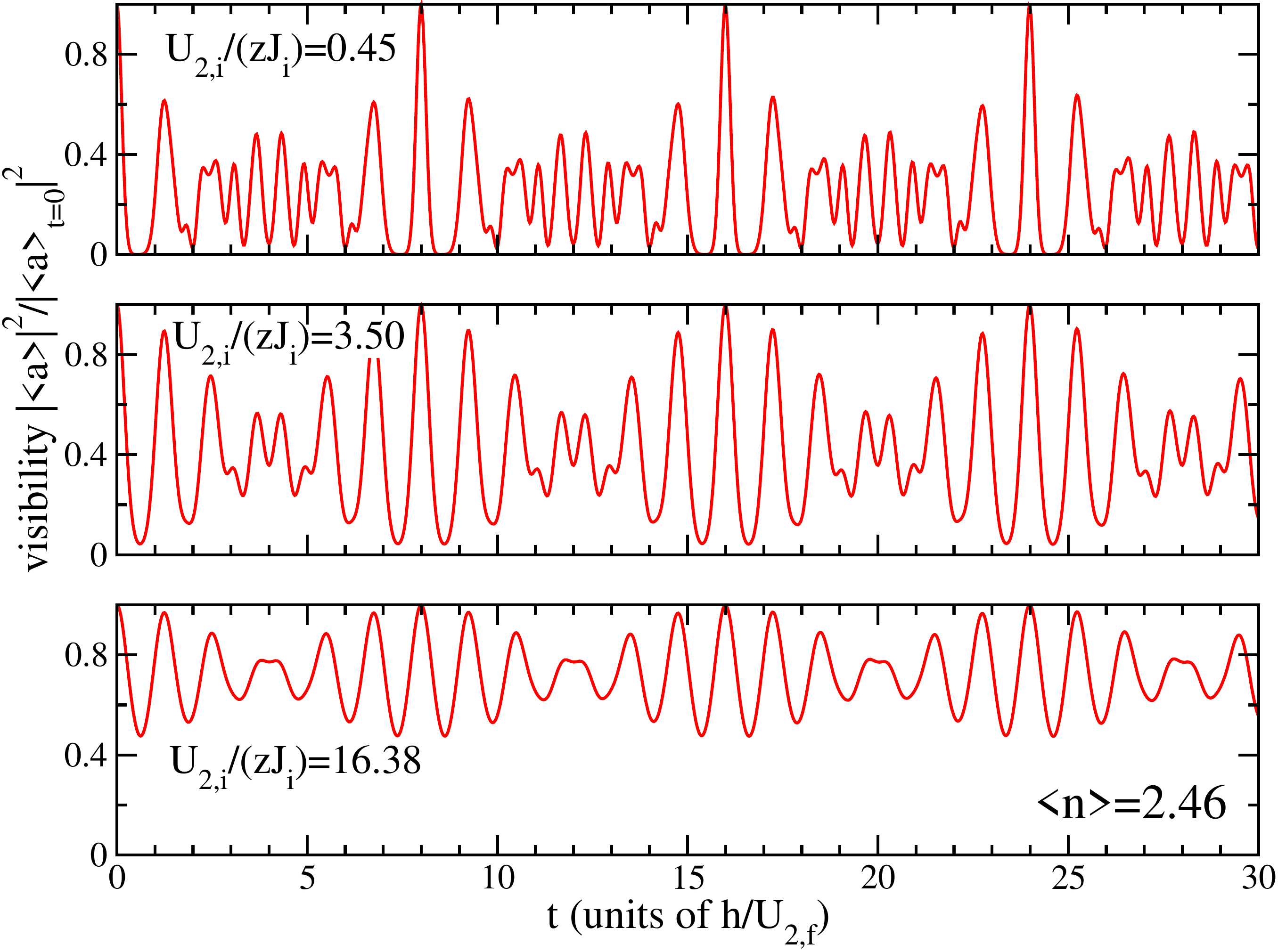}
\caption{(Color online) Visibility $|\langle a \rangle|^{2}/|\langle a
\rangle_{t=0}|^{2}$ as a function of hold time $t$ after a sudden increase
of lattice depth. The three panels correspond to three initial states
characterized by the ratio $\protect\xi=U_{2,i}/(zJ_{i})$ before the
increase. The mean atom number is $\langle n\rangle=2.46$ and the
(renormalized) pair-wise interaction strength $U_{2,f}=0.0928\hbar\protect%
\omega_{f}$ for all traces. Time is in units of $h/U_{2,f}$ (e.g., $0.2$ ms
for $^{87}$Rb in a 41 $E_R$ deep lattice.)}
\label{timetrace}
\end{figure}

\section{Interference patterns}

After hold time $t$, the optical lattice is turned off and the atoms freely
expand. The visibility of the spatial interference pattern versus $t$ yields
information about both the initial state $|\Psi _{i}\rangle $ and the
dynamics. In \cite{Greiner2002b} it was shown that the visibility is $%
|\langle a\rangle |^{2}/|\langle a\rangle _{t=0}|^{2}=|\langle a\rangle
|^{2}/\phi ^{2},$ where the expectation value is over $|\Psi _{f}(t)\rangle $%
. A zero visibility corresponds to a Gaussian spatial density profile, while
a value of one corresponds to a diffraction pattern with maximum contrast.

Figure \ref{timetrace} shows time traces of the visibility versus $t$ for
three initial states differing in the value of $\xi =U_{2,i}/(zJ_{i})$, and $%
\langle n\rangle =2.46$ as in Fig.~\ref{initial}. The value $%
U_{2,f}=0.0928\hbar \omega _{f}$ is typical for experiments with $^{87}$Rb
atoms in \textquotedblleft deep\textquotedblright\ optical lattices. The
traces show fast oscillations with timescale~$\sim h/U_{2,f}$, and a slower
envelope with a timescale $\sim h/U_{3,f}$. The pattern is more complex for
initial states close to an ideal coherent superfluid, and there are near
total collapses of the matter-wave coherence at regular intervals. For
increasing $\xi $ the initial state is increasingly number squeezed, and the
traces become smoother. In qualitative agreement with the data in~\cite%
{Will2010}, there are only partial collapses of the visibilities.

The value for $\langle n \rangle$ is chosen to accentuate the effect of both
two- and three-body interactions and corresponds to easily obtained
experimental parameters. For smaller $\langle n \rangle$ we find that the
contrast and three-body effects are reduced, while for larger $\langle n
\rangle$ the effective Hamiltonian needs to be improved.

\begin{figure}[t]
\includegraphics[width=8.5cm]{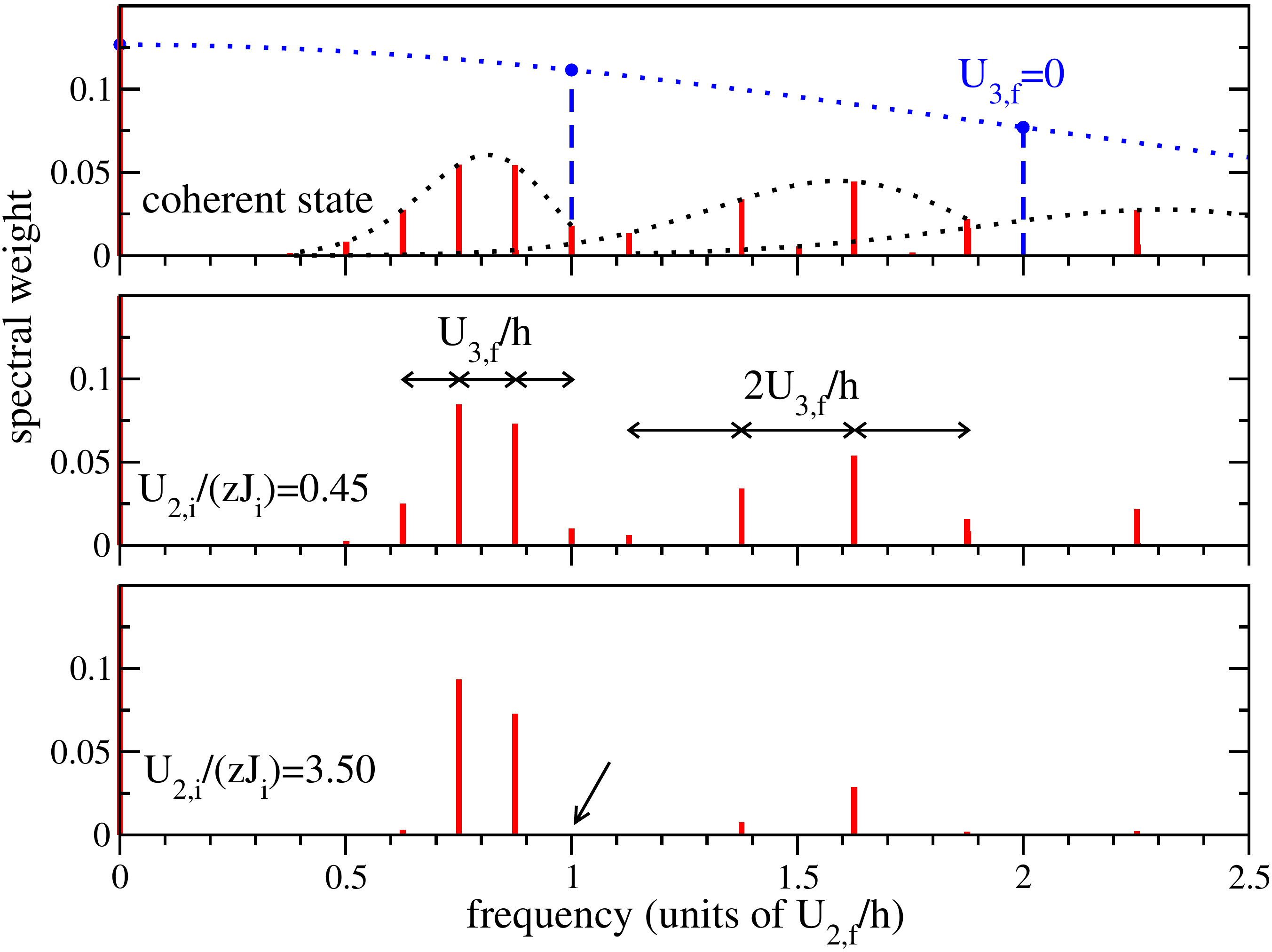}
\caption{(Color online) Spectrum of visibility as function of positive
frequency $\protect\nu$ for initial coherent state (top panel) and two
initial states obtained using mean-field decoupling approximation (bottom
two panels). Time traces of the latter two spectra are shown in Fig.~\protect
\ref{timetrace}. We use $\langle n\rangle=2.46$ and $U_{2,f}=0.0928\hbar%
\protect\omega_{f}$. Spectral lines correspond to the vertical lines. The
spacings between dominant lines are shown in middle panel. The $\protect\nu%
=0 $ line strength increases from 0.2 to 0.3 from the top to bottom panel.
The arrow in the bottom panel shows the vanishing of a line at $h\protect\nu%
=U_{2,f}$ for large ratio $U_{2,i}/(zJ_{i})$. For coherent states, the
spectrum with and without effective three-body interaction are shown as full
and dashed lines, respectively. For $U_{3,f}=0$ only frequencies with
integer multiples of $U_{2,f}/h$ have nonzero line strength. The dotted
curves are analytical expressions for the line strength, discussed in the
text. }
\label{spectrum}
\end{figure}

The behavior of the visibility in Fig.~\ref{timetrace} can be elucidated by
performing a Fourier expansion. The allowed frequencies $\nu $ are
differences between frequencies $E_{n}/h$ governing the evolution of $|\Psi
_{f}(t)\rangle $ \cite{Will2010}. Spectra of the visibility are shown in
Fig.~\ref{spectrum}. The top panel shows the spectrum for an initial, ideal
coherent state, while the bottom two panels show spectra for states
calculated with the decoupling approximation. Consistent with the smoother
traces in Fig.~\ref{timetrace}, fewer features are visible when $\xi $
increases. For example, the arrow in the bottom panel of Fig.~\ref{spectrum}
highlights the disappearance of the line at $U_{2,f}/h$. This behavior is
not obvious from examination of $H_{f}$ alone and dramatically illustrates
why the analysis presented here of the dynamics plus initial state is
required to fully understand collapse and revival spectra.

The frequency components of $|\langle a\rangle|^{2}$ follow a regular
pattern, yielding a method for obtaining the two- and three-body coupling
strengths from the dynamics. A progression of equally spaced frequencies
ends on or near each of the limits $kU_{2,f}/h$ for positive integer $k$.
For example, for $\nu\leq U_{2,f}/h$ features appear at 

\begin{equation}
U_{2,f}/h+mU_{3,f}/h,  
\end{equation}%
for $m=0,1,2,\cdots.$ In fact, for $\nu\leq
kU_{2,f}/h $ and $k\geq1$ the progression is $%
kU_{2,f}/h+(mk+k(k-1)/2)U_{3,f}/h$ for $m=0,1,2,\cdots$ corresponding to
spacings between lines of $kU_{3,f}/h$. There are no lines at frequencies $%
kU_{2,f}/h$ except for $k=1$.

The Fock state amplitudes $c_{n} $ of the initial state $|\Psi _{i}\rangle$
can, in turn, be directly obtained from the line strengths. For example, the
lines at $U_{2,f}/h+mU_{3,f}/h$ have a strength of $\sqrt{(m+2)(m+1)}
|c_{m+2}^{\ast}c_{m+1}^{2}c_{m}^{\ast}|/\phi^{2}$ (the expansion was
previously shown in Ref.~\cite{Will2010}), which depends on amplitudes of
three Fock states. Again illustrating that complete analysis of the spectra
is subtle, we note that for $m=0$ the strength of the $U_{2,f}/h$ spectral
line is proportional to $c_{0}$, the amplitude in the zero atom state! For
example, for coherent or small $\xi$ initial states the zero atom component
has significant population (see Fig.~\ref{initial}) and the line at $%
U_{2,f}/h $ is visible; while for large $\xi$ it disappears. The variations
in line strength therefore correspond to the degree of superfluidity or
number squeezing in the initial lattice state.

For coherent states, closed-form expressions for the line strengths may be
obtained allowing a detailed comparison between the collapse-and-revival
spectra for coherent and more realistic squeezed states. For a coherent
state and $U_{3,f}=0$, the visibility is $e^{2\phi^{2}(\cos(U_{2,f}t/%
\hbar)-1)}$ \cite{Will2010}, which in the frequency domain implies lines at $%
kU_{2,f}/h$ for integer $k$ with strength $e^{-2\phi^{2}}I_{k}(2\phi^{2})$,
where $I_{k}(z) $ is a modified Bessel function. This spectrum is shown in
the top panel of Fig.~\ref{spectrum}, where the dotted curve connecting the
blue vertical lines is found by replacing the integer order $k$ of the
Bessel function by $h\nu/U_{2,f}$.

The spectrum for an initial coherent state with $U_{3,f}\neq 0$ can also be
solved analytically. Frequencies $U_{2,f}/h+mU_{3,f}/h$ have line strengths 
\begin{equation}
\phi ^{2}e^{-2\phi ^{2}}\phi ^{4m}/(\Gamma (m+2)\Gamma (m+1)),
\end{equation}
and frequencies $2U_{2,f}/h+(2m+1)U_{3,f}/h$ have line strengths 
\begin{equation}
\phi ^{4}e^{-2\phi ^{2}}\phi ^{4m}/(\Gamma (m+3)\Gamma (m+1)),
\end{equation}
where $\Gamma (z)$ is the Gamma function. Expressions for frequencies below $%
kU_{2,f}/h$ for $k>2$ can also be found. This spectrum is shown in the top
panel of Fig.~\ref{spectrum}, where the dotted curve connecting the lines is
found by replacing $m$ by the associated function of $\nu $. For frequencies
below $U_{2}/h$ only lines near $U_{2,f}/h+\langle n\rangle U_{3,f}/h$
survive and, similarly, only frequencies near $2U_{2,f}/h+(2\langle n\rangle
+1)U_{3,f}/h$ survive.

\begin{figure}[tbp]
\includegraphics[width=8.5cm]{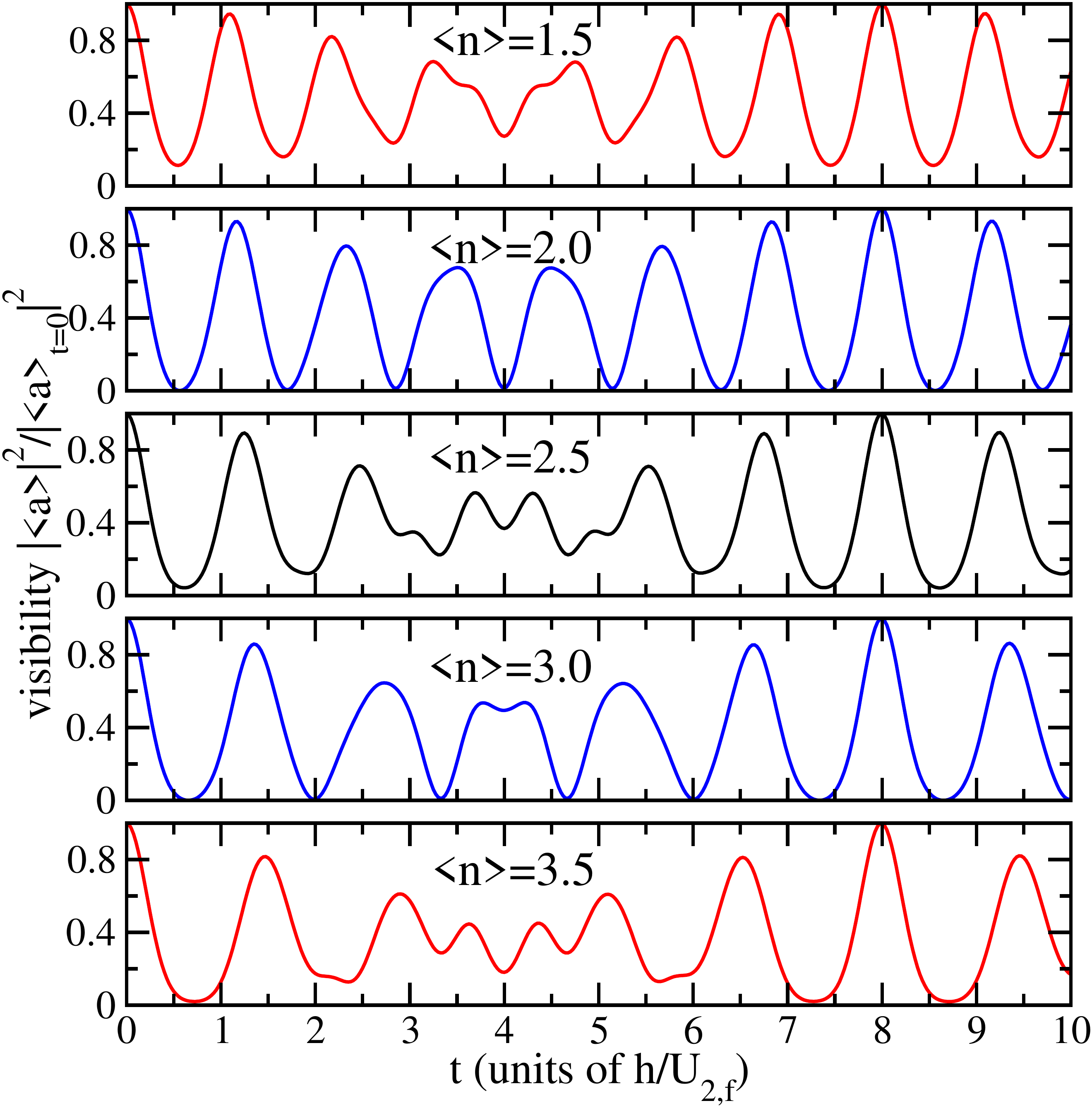}
\caption{(Color Online) Visibility as a function of hold time for initial
states with a mean atom number $\langle n\rangle$ of 1.5, 2.0, 2.5, 3.0, and
3.5 going from the top to bottom panel. We use $\protect\xi%
=U_{2,i}/(zJ_i)=3.50$ and $U_{2,f}=0.0928\hbar\protect\omega_f$. The initial
number probability and time trace for $\langle n\rangle=2.5$ closely
resembles that for $\langle n\rangle=2.46$ and $\protect\xi=3.50$ shown in
Figs.~\protect\ref{initial} and \protect\ref{timetrace}, respectively. }
\label{tracevsN}
\end{figure}

We now study the dependence of the collapse and revival experiments on
initial mean atom number, or equivalently the chemical potential. Figure \ref%
{tracevsN} shows the visibility as a function of hold time for several
initial values of mean atom number. The system parameters $J_{i}$ and $%
U_{2,i}$ are held fixed and chosen such that the initial superfluid state
shows significant squeezing. In contrast with Fig.~\ref{timetrace} where the
patterns change dramatically with $\xi $, Fig.~\ref{tracevsN} shows
oscillation patterns that are robust to changes in mean atom number. This
robustness is because the frequencies in the time-traces do not depend on
the average atom number, at least in the regime where the effective
Hamiltonian $H_{f}$ is valid. For example, the fast oscillations related to $%
h/U_{2,f}$ are in phase and the contrast is nearly independent of $\langle
n\rangle $. This behavior can be understood from our analysis of the
frequency components, which are independent of $c_{n}$ and thus also
mean-atom number. (For example, for $\nu \leq U_{2,f}/h$ spectral lines
appear at $U_{2,f}/h+mU_{3,f}/h$ for $m=0,1,2,\cdots $.)

In Fig.~\ref{tracevsN} the differences in the time evolution are only
noticeable at half integer multiples of $h/U_{3,f}$, corresponding to $%
t\approx 4.0 h/U_{2,f}$ in the figure. At these times the contribution to
the phase factor $e^{-iE_nt/\hbar}$ appearing in the evolution of $%
|\Psi_f(t)\rangle$ due to the three-body interaction is $-1$ for the $n =
3,7,11,...$ Fock states $|n\rangle$.

\section{Mean atom-number uncertainties}

\begin{figure}[b]
\includegraphics[width=8.5cm]{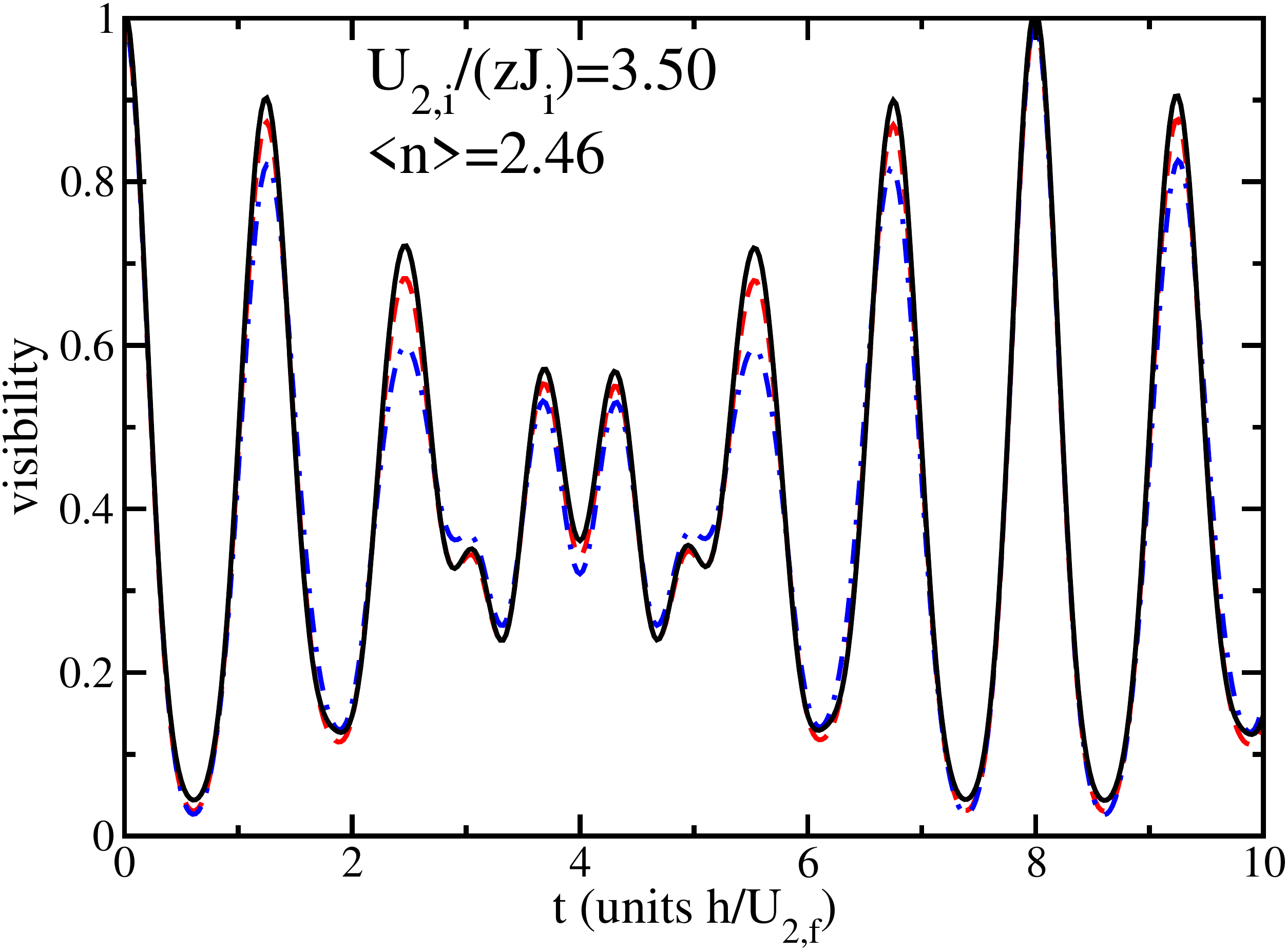}
\caption{(Color online) Visibility as a function of hold time for three
different averages over an experimental uncertainty in initial mean-atom
number (or equivalently chemical potential) in the superfluid. We use $%
\protect\xi=U_{2,i}/(zJ_i)=3.50$ and $U_{2,f}=0.0928\hbar\protect\omega_f$.
The full (black) curve, already shown in Fig.~\protect\ref{timetrace},
assumes complete reproducibility at $\langle n\rangle=2.46$, while the
dashed (red) curve use a Gaussian distribution with a mean atom number of
2.46 and a standard deviation equal to 0.25 (or relative uncertainty of
10\%). The dash-dotted (blue) curve is obtained for a standard deviation of
0.5 (or relative uncertainty of 20\%).}
\label{average}
\end{figure}

We can quantify possible dephasing effects due to experimental uncertainties
in preparing the superfluid with the same mean atom number $\langle n\rangle 
$ or equivalently the same chemical potential. An experimental sequence
requires a new preparation of the superfluid for each hold time as the
measurement of the interference pattern is destructive. We model this
uncertainty by introducing a probability distribution $p(\langle n\rangle )$
over $\langle n\rangle $ and calculate 
\begin{equation}
\rho (t)=\int_{0}^{\infty }d\langle n\rangle \,p(\langle n\rangle
)\,|\langle a\rangle |^{2}\,,
\end{equation}%
which is time dependent through $\langle a\rangle $. The visibility becomes $%
\rho (t)/\rho (0)$. Figure \ref{average} shows time traces of the visibility
after averaging over a Gaussian distributed $p(\langle n\rangle )$
normalized to one for $\langle n\rangle \geq 0$. The three curves correspond
to Gaussian distributions centered at an average of $2.46$ atoms, and with a
square root of the variance (standard deviation) that is either zero, 10\%,
or 20\% of the average value, respectively. As in Fig.~\ref{tracevsN} we
immediately note that the revivals occur at the same times with only small
changes in amplitude, demonstrating that collapse and revival experiments
are robust under small variations in mean-atom number. A consequence of this
independence is that in the frequency domain there is no broadening of the
spectral lines due to uncertainties in the mean atom number after averaging over experimental realizations. There is broadening due to other mechanisms.
For example, inhomogeneities leading to a variation of  $U_{2,f}$ across the lattice, as well as thermal effects, will cause both broadening and dephasing. The successes of the experiments in Ref.~\cite{Will2010} are made possible by the ability to reduce these effects sufficiently to allow observation of tens of oscillations.

\section{Conclusions}

We have shown that collapse-and-revival dynamics of atoms in optical
lattices are strongly sensitive to both the initial many-body ground state
and effective higher-body dynamics when projected into a deep lattice. The
spectral analysis of the time evolution gives detailed information about
effective interaction strengths as well as amplitudes of atom-number states
superposed in the initial state. For example, number squeezing is apparent
in the bottom two panels of Fig.~\ref{spectrum}, where fewer frequency
components are present than for the coherent state shown in the top panel.
We also found that the frequencies follow regular patterns that are independent of the initial superposition of number states, and the line strengths give information about the initial state.

Our analysis is based on various approximations. Future work should evaluate
corrections to the decoupling approximation for small lattice depths. For
example, Ref.~\cite{Freericks1994} showed a marked effect of dimensionality
on the Mott phase transition, and Ref.~\cite{Wolf2010} analyzed the role of
tunneling in collapse and revival dynamics. Finite temperature effects may
also be important. Wedding-cake structures~\cite{Jaksch1998} that occur with
additional weak confinement can be included using a local-density
approximation. Three-body recombination should be included in the analysis
when in the future the dynamics are studied over a longer period of time than currently observed. The inclusion of the anharmonicity of the
lattice-site potential, which reduces the level spacing and affects the
relationship between the effective two- and three-body interaction, can be
taken into account to construct an improved theoretical model. 

Despite these technical limitations, however, the combination of effects considered here captures important features of the collapse-and-revival dynamics. By treating the interaction strengths $U_{2,f}$ and $U_{3,f}$, and the initial state coefficients $c_n$, as free parameters in a fit to the experimental data it should be possible to both test our model, and correct for some of its limitations, for example, the neglect of the effects of anharmonicities. Inconsistencies in the fit could then provide insight into the breakdown of the mean-field theory (including thermal effects) for the initial state or the role of nonperturbative effects in the final state dynamics.

\bibliography{refs}

\begin{thebibliography}{29}
\expandafter\ifx\csname natexlab\endcsname\relax\def\natexlab#1{#1}\fi
\expandafter\ifx\csname bibnamefont\endcsname\relax
  \def\bibnamefont#1{#1}\fi
\expandafter\ifx\csname bibfnamefont\endcsname\relax
  \def\bibfnamefont#1{#1}\fi
\expandafter\ifx\csname citenamefont\endcsname\relax
  \def\citenamefont#1{#1}\fi
\expandafter\ifx\csname url\endcsname\relax
  \def\url#1{\texttt{#1}}\fi
\expandafter\ifx\csname urlprefix\endcsname\relax\def\urlprefix{URL }\fi
\providecommand{\bibinfo}[2]{#2}
\providecommand{\eprint}[2][]{\url{#2}}

\bibitem[{\citenamefont{Morsch and Oberthaler}(2006)}]{Morsch2006}
\bibinfo{author}{\bibfnamefont{O.}~\bibnamefont{Morsch}} \bibnamefont{and}
  \bibinfo{author}{\bibfnamefont{M.}~\bibnamefont{Oberthaler}},
  \bibinfo{journal}{Rev. Mod. Phys.} \textbf{\bibinfo{volume}{78}},
  \bibinfo{pages}{179} (\bibinfo{year}{2006}).

\bibitem[{\citenamefont{Bloch et~al.}(2008)\citenamefont{Bloch, Dalibard, and
  Zwerger}}]{Bloch2008}
\bibinfo{author}{\bibfnamefont{I.}~\bibnamefont{Bloch}},
  \bibinfo{author}{\bibfnamefont{J.}~\bibnamefont{Dalibard}}, \bibnamefont{and}
  \bibinfo{author}{\bibfnamefont{W.}~\bibnamefont{Zwerger}},
  \bibinfo{journal}{Rev. Mod. Phys.} \textbf{\bibinfo{volume}{80}},
  \bibinfo{pages}{885} (\bibinfo{year}{2008}).

\bibitem[{\citenamefont{Jaksch et~al.}(1998)\citenamefont{Jaksch, Bruder,
  Cirac, Gardiner, and Zoller}}]{Jaksch1998}
\bibinfo{author}{\bibfnamefont{D.}~\bibnamefont{Jaksch}},
  \bibinfo{author}{\bibfnamefont{C.}~\bibnamefont{Bruder}},
  \bibinfo{author}{\bibfnamefont{J.~I.} \bibnamefont{Cirac}},
  \bibinfo{author}{\bibfnamefont{C.~W.} \bibnamefont{Gardiner}},
  \bibnamefont{and} \bibinfo{author}{\bibfnamefont{P.}~\bibnamefont{Zoller}},
  \bibinfo{journal}{Phys. Rev. Lett.} \textbf{\bibinfo{volume}{81}},
  \bibinfo{pages}{3108} (\bibinfo{year}{1998}).

\bibitem[{\citenamefont{Fisher et~al.}(1989)\citenamefont{Fisher, Weichman,
  Grinstein, and Fisher}}]{Fisher1989}
\bibinfo{author}{\bibfnamefont{M.~P.~A.} \bibnamefont{Fisher}},
  \bibinfo{author}{\bibfnamefont{P.~B.} \bibnamefont{Weichman}},
  \bibinfo{author}{\bibfnamefont{G.}~\bibnamefont{Grinstein}},
  \bibnamefont{and} \bibinfo{author}{\bibfnamefont{D.~S.}
  \bibnamefont{Fisher}}, \bibinfo{journal}{Phys. Rev. B}
  \textbf{\bibinfo{volume}{40}}, \bibinfo{pages}{546} (\bibinfo{year}{1989}).

\bibitem[{\citenamefont{Greiner
  et~al.}(2002{\natexlab{a}})\citenamefont{Greiner, Mandel, Esslinger,
  H\"ansch, and Bloch}}]{Greiner2002a}
\bibinfo{author}{\bibfnamefont{M.}~\bibnamefont{Greiner}},
  \bibinfo{author}{\bibfnamefont{O.}~\bibnamefont{Mandel}},
  \bibinfo{author}{\bibfnamefont{T.}~\bibnamefont{Esslinger}},
  \bibinfo{author}{\bibfnamefont{T.~W.} \bibnamefont{H\"ansch}},
  \bibnamefont{and} \bibinfo{author}{\bibfnamefont{I.}~\bibnamefont{Bloch}},
  \bibinfo{journal}{Nature} \textbf{\bibinfo{volume}{415}}, \bibinfo{pages}{39}
  (\bibinfo{year}{2002}{\natexlab{a}}).

\bibitem[{\citenamefont{Nelson et~al.}(2007)\citenamefont{Nelson, Li, and
  Weiss}}]{Nelson2007}
\bibinfo{author}{\bibfnamefont{K.~D.} \bibnamefont{Nelson}},
  \bibinfo{author}{\bibfnamefont{X.}~\bibnamefont{Li}}, \bibnamefont{and}
  \bibinfo{author}{\bibfnamefont{D.~S.} \bibnamefont{Weiss}},
  \bibinfo{journal}{Nat. Phys.} \textbf{\bibinfo{volume}{3}},
  \bibinfo{pages}{556} (\bibinfo{year}{2007}).

\bibitem[{\citenamefont{Gemelke et~al.}(2009)\citenamefont{Gemelke, Zhang,
  Hung, and Chin}}]{Gemelke2009}
\bibinfo{author}{\bibfnamefont{N.}~\bibnamefont{Gemelke}},
  \bibinfo{author}{\bibfnamefont{X.}~\bibnamefont{Zhang}},
  \bibinfo{author}{\bibfnamefont{C.}~\bibnamefont{Hung}}, \bibnamefont{and}
  \bibinfo{author}{\bibfnamefont{C.}~\bibnamefont{Chin}},
  \bibinfo{journal}{Nature} \textbf{\bibinfo{volume}{460}},
  \bibinfo{pages}{995} (\bibinfo{year}{2009}).

\bibitem[{\citenamefont{Bakr et~al.}(2010)\citenamefont{Bakr, Peng, Tai, Ma,
  Simon, Gillen, F�lling, Pollet, and Greiner}}]{Bakr2010}
\bibinfo{author}{\bibfnamefont{W.~S.} \bibnamefont{Bakr}},
  \bibinfo{author}{\bibfnamefont{A.}~\bibnamefont{Peng}},
  \bibinfo{author}{\bibfnamefont{M.~E.} \bibnamefont{Tai}},
  \bibinfo{author}{\bibfnamefont{R.}~\bibnamefont{Ma}},
  \bibinfo{author}{\bibfnamefont{J.}~\bibnamefont{Simon}},
  \bibinfo{author}{\bibfnamefont{J.~I.} \bibnamefont{Gillen}},
  \bibinfo{author}{\bibfnamefont{S.}~\bibnamefont{F�lling}},
  \bibinfo{author}{\bibfnamefont{L.}~\bibnamefont{Pollet}}, \bibnamefont{and}
  \bibinfo{author}{\bibfnamefont{M.}~\bibnamefont{Greiner}},
  \bibinfo{journal}{Science} \textbf{\bibinfo{volume}{329}},
  \bibinfo{pages}{547} (\bibinfo{year}{2010}).

\bibitem[{\citenamefont{Jordens et~al.}(2008)\citenamefont{Jordens, Strohmaier,
  Gunter, Moritz, and Esslinger}}]{Jordens2008}
\bibinfo{author}{\bibfnamefont{R.}~\bibnamefont{Jordens}},
  \bibinfo{author}{\bibfnamefont{N.}~\bibnamefont{Strohmaier}},
  \bibinfo{author}{\bibfnamefont{K.}~\bibnamefont{Gunter}},
  \bibinfo{author}{\bibfnamefont{H.}~\bibnamefont{Moritz}}, \bibnamefont{and}
  \bibinfo{author}{\bibfnamefont{T.}~\bibnamefont{Esslinger}},
  \bibinfo{journal}{Nature} \textbf{\bibinfo{volume}{455}},
  \bibinfo{pages}{204} (\bibinfo{year}{2008}).

\bibitem[{\citenamefont{Hung et~al.}(2010)\citenamefont{Hung, Zhang, Gemelke,
  and Chin}}]{Hung2010}
\bibinfo{author}{\bibfnamefont{C.}~\bibnamefont{Hung}},
  \bibinfo{author}{\bibfnamefont{X.}~\bibnamefont{Zhang}},
  \bibinfo{author}{\bibfnamefont{N.}~\bibnamefont{Gemelke}}, \bibnamefont{and}
  \bibinfo{author}{\bibfnamefont{C.}~\bibnamefont{Chin}},
  \bibinfo{journal}{Phys. Rev. Lett.} \textbf{\bibinfo{volume}{104}},
  \bibinfo{pages}{160403} (\bibinfo{year}{2010}).

\bibitem[{\citenamefont{Ho and Zhou}(2010)}]{Ho2010}
\bibinfo{author}{\bibfnamefont{T.}~\bibnamefont{Ho}} \bibnamefont{and}
  \bibinfo{author}{\bibfnamefont{Q.}~\bibnamefont{Zhou}},
  \bibinfo{journal}{Nat. Phys.} \textbf{\bibinfo{volume}{6}},
  \bibinfo{pages}{131} (\bibinfo{year}{2010}).

\bibitem[{\citenamefont{Moritz et~al.}(2005)\citenamefont{Moritz, St\"oferle,
  G\"unter, K\"ohl, and Esslinger}}]{Moritz2005}
\bibinfo{author}{\bibfnamefont{H.}~\bibnamefont{Moritz}},
  \bibinfo{author}{\bibfnamefont{T.}~\bibnamefont{St\"oferle}},
  \bibinfo{author}{\bibfnamefont{K.}~\bibnamefont{G\"unter}},
  \bibinfo{author}{\bibfnamefont{M.}~\bibnamefont{K\"ohl}}, \bibnamefont{and}
  \bibinfo{author}{\bibfnamefont{T.}~\bibnamefont{Esslinger}},
  \bibinfo{journal}{Phys. Rev. Lett.} \textbf{\bibinfo{volume}{94}},
  \bibinfo{pages}{210401} (\bibinfo{year}{2005}).

\bibitem[{\citenamefont{Ospelkaus et~al.}(2006)\citenamefont{Ospelkaus,
  Ospelkaus, Humbert, Ernst, Sengstock, and Bongs}}]{Ospelkaus2006}
\bibinfo{author}{\bibfnamefont{C.}~\bibnamefont{Ospelkaus}},
  \bibinfo{author}{\bibfnamefont{S.}~\bibnamefont{Ospelkaus}},
  \bibinfo{author}{\bibfnamefont{L.}~\bibnamefont{Humbert}},
  \bibinfo{author}{\bibfnamefont{P.}~\bibnamefont{Ernst}},
  \bibinfo{author}{\bibfnamefont{K.}~\bibnamefont{Sengstock}},
  \bibnamefont{and} \bibinfo{author}{\bibfnamefont{K.}~\bibnamefont{Bongs}},
  \bibinfo{journal}{Phys. Rev. Lett.} \textbf{\bibinfo{volume}{97}},
  \bibinfo{pages}{120402} (\bibinfo{year}{2006}).

\bibitem[{\citenamefont{Widera et~al.}(2006)\citenamefont{Widera, Gerbier,
  F\"olling, Gericke, Mandel, and Bloch}}]{Widera2006}
\bibinfo{author}{\bibfnamefont{A.}~\bibnamefont{Widera}},
  \bibinfo{author}{\bibfnamefont{F.}~\bibnamefont{Gerbier}},
  \bibinfo{author}{\bibfnamefont{S.}~\bibnamefont{F\"olling}},
  \bibinfo{author}{\bibfnamefont{T.}~\bibnamefont{Gericke}},
  \bibinfo{author}{\bibfnamefont{O.}~\bibnamefont{Mandel}}, \bibnamefont{and}
  \bibinfo{author}{\bibfnamefont{I.}~\bibnamefont{Bloch}},
  \bibinfo{journal}{New J. Phys.} \textbf{\bibinfo{volume}{8}},
  \bibinfo{pages}{152} (\bibinfo{year}{2006}).

\bibitem[{\citenamefont{Greiner
  et~al.}(2002{\natexlab{b}})\citenamefont{Greiner, Mandel, H\"ansch, and
  Bloch}}]{Greiner2002b}
\bibinfo{author}{\bibfnamefont{M.}~\bibnamefont{Greiner}},
  \bibinfo{author}{\bibfnamefont{O.}~\bibnamefont{Mandel}},
  \bibinfo{author}{\bibfnamefont{T.~W.} \bibnamefont{H\"ansch}},
  \bibnamefont{and} \bibinfo{author}{\bibfnamefont{I.}~\bibnamefont{Bloch}},
  \bibinfo{journal}{Nature} \textbf{\bibinfo{volume}{419}},
  \bibinfo{pages}{51�54} (\bibinfo{year}{2002}{\natexlab{b}}).

\bibitem[{\citenamefont{{Sebby-Strabley}
  et~al.}(2007)\citenamefont{{Sebby-Strabley}, Brown, Anderlini, Lee, Phillips,
  Porto, and Johnson}}]{Sebby2007}
\bibinfo{author}{\bibfnamefont{J.}~\bibnamefont{{Sebby-Strabley}}},
  \bibinfo{author}{\bibfnamefont{B.~L.} \bibnamefont{Brown}},
  \bibinfo{author}{\bibfnamefont{M.}~\bibnamefont{Anderlini}},
  \bibinfo{author}{\bibfnamefont{P.~J.} \bibnamefont{Lee}},
  \bibinfo{author}{\bibfnamefont{W.~D.} \bibnamefont{Phillips}},
  \bibinfo{author}{\bibfnamefont{J.~V.} \bibnamefont{Porto}}, \bibnamefont{and}
  \bibinfo{author}{\bibfnamefont{P.~R.} \bibnamefont{Johnson}},
  \bibinfo{journal}{Phys. Rev. Lett.} \textbf{\bibinfo{volume}{98}},
  \bibinfo{pages}{200405} (\bibinfo{year}{2007}).

\bibitem[{\citenamefont{Will et~al.}(2010)\citenamefont{Will, Best, Schneider,
  Hackermuller, Luhmann, and Bloch}}]{Will2010}
\bibinfo{author}{\bibfnamefont{S.}~\bibnamefont{Will}},
  \bibinfo{author}{\bibfnamefont{T.}~\bibnamefont{Best}},
  \bibinfo{author}{\bibfnamefont{U.}~\bibnamefont{Schneider}},
  \bibinfo{author}{\bibfnamefont{L.}~\bibnamefont{Hackermuller}},
  \bibinfo{author}{\bibfnamefont{D.}~\bibnamefont{Luhmann}}, \bibnamefont{and}
  \bibinfo{author}{\bibfnamefont{I.}~\bibnamefont{Bloch}},
  \bibinfo{journal}{Nature} \textbf{\bibinfo{volume}{465}},
  \bibinfo{pages}{197} (\bibinfo{year}{2010}).

\bibitem[{\citenamefont{Toth et~al.}(2008)\citenamefont{Toth, Rey, and
  Blakie}}]{Toth2008}
\bibinfo{author}{\bibfnamefont{E.}~\bibnamefont{Toth}},
  \bibinfo{author}{\bibfnamefont{A.~M.} \bibnamefont{Rey}}, \bibnamefont{and}
  \bibinfo{author}{\bibfnamefont{P.~B.} \bibnamefont{Blakie}},
  \bibinfo{journal}{Phys. Rev. A} \textbf{\bibinfo{volume}{78}},
  \bibinfo{pages}{013627} (\bibinfo{year}{2008}).

\bibitem[{\citenamefont{Johnson et~al.}(2009)\citenamefont{Johnson, Tiesinga,
  Porto, and Williams}}]{Johnson2009}
\bibinfo{author}{\bibfnamefont{P.~R.} \bibnamefont{Johnson}},
  \bibinfo{author}{\bibfnamefont{E.}~\bibnamefont{Tiesinga}},
  \bibinfo{author}{\bibfnamefont{J.~V.} \bibnamefont{Porto}}, \bibnamefont{and}
  \bibinfo{author}{\bibfnamefont{C.~J.} \bibnamefont{Williams}},
  \bibinfo{journal}{New J. Phys.} \textbf{\bibinfo{volume}{11}},
  \bibinfo{pages}{093022} (\bibinfo{year}{2009}).

\bibitem[{\citenamefont{Schachenmayer et~al.}(2011)\citenamefont{Schachenmayer,
  Daley, and Zoller}}]{Schachenmayer2011}
\bibinfo{author}{\bibfnamefont{J.}~\bibnamefont{Schachenmayer}},
  \bibinfo{author}{\bibfnamefont{A.~J.} \bibnamefont{Daley}}, \bibnamefont{and}
  \bibinfo{author}{\bibfnamefont{P.}~\bibnamefont{Zoller}},
  \bibinfo{journal}{arxiv 1101.2385}  (\bibinfo{year}{2011}).

\bibitem[{\citenamefont{Sheshadri et~al.}(1993)\citenamefont{Sheshadri,
  Krishnamurthy, Pandit, and Ramakrishnan}}]{Sheshadri1993}
\bibinfo{author}{\bibfnamefont{K.}~\bibnamefont{Sheshadri}},
  \bibinfo{author}{\bibfnamefont{H.~R.} \bibnamefont{Krishnamurthy}},
  \bibinfo{author}{\bibfnamefont{R.}~\bibnamefont{Pandit}}, \bibnamefont{and}
  \bibinfo{author}{\bibfnamefont{T.~V.} \bibnamefont{Ramakrishnan}},
  \bibinfo{journal}{Europhys. Lett.} \textbf{\bibinfo{volume}{22}},
  \bibinfo{pages}{257} (\bibinfo{year}{1993}).

\bibitem[{\citenamefont{van Oosten et~al.}(2001)\citenamefont{van Oosten,
  van~der Straten, and Stoof}}]{vanOosten2001}
\bibinfo{author}{\bibfnamefont{D.}~\bibnamefont{van Oosten}},
  \bibinfo{author}{\bibfnamefont{P.}~\bibnamefont{van~der Straten}},
  \bibnamefont{and} \bibinfo{author}{\bibfnamefont{H.~T.~C.}
  \bibnamefont{Stoof}}, \bibinfo{journal}{Phys. Rev. A}
  \textbf{\bibinfo{volume}{63}}, \bibinfo{pages}{053601}
  (\bibinfo{year}{2001}).

\bibitem[{\citenamefont{Freericks and Monien}(1994)}]{Freericks1994}
\bibinfo{author}{\bibfnamefont{J.~K.} \bibnamefont{Freericks}}
  \bibnamefont{and} \bibinfo{author}{\bibfnamefont{H.}~\bibnamefont{Monien}},
  \bibinfo{journal}{Europhys. Lett.} \textbf{\bibinfo{volume}{26}},
  \bibinfo{pages}{545} (\bibinfo{year}{1994}).

\bibitem[{\citenamefont{Huang and Yang}(1957)}]{Huang1957}
\bibinfo{author}{\bibfnamefont{K.}~\bibnamefont{Huang}} \bibnamefont{and}
  \bibinfo{author}{\bibfnamefont{C.~N.} \bibnamefont{Yang}},
  \bibinfo{journal}{Phys. Rev.} \textbf{\bibinfo{volume}{105}},
  \bibinfo{pages}{767} (\bibinfo{year}{1957}).

\bibitem[{\citenamefont{Chin et~al.}(2010)\citenamefont{Chin, Grimm, Julienne,
  and Tiesinga}}]{Chin2010}
\bibinfo{author}{\bibfnamefont{C.}~\bibnamefont{Chin}},
  \bibinfo{author}{\bibfnamefont{R.}~\bibnamefont{Grimm}},
  \bibinfo{author}{\bibfnamefont{P.}~\bibnamefont{Julienne}}, \bibnamefont{and}
  \bibinfo{author}{\bibfnamefont{E.}~\bibnamefont{Tiesinga}},
  \bibinfo{journal}{Rev. Mod. Phys.} \textbf{\bibinfo{volume}{82}},
  \bibinfo{pages}{1225} (\bibinfo{year}{2010}).

\bibitem[{\citenamefont{Wolf et~al.}(2010)\citenamefont{Wolf, Hen, and
  Rigol}}]{Wolf2010}
\bibinfo{author}{\bibfnamefont{F.~A.} \bibnamefont{Wolf}},
  \bibinfo{author}{\bibfnamefont{I.}~\bibnamefont{Hen}}, \bibnamefont{and}
  \bibinfo{author}{\bibfnamefont{M.}~\bibnamefont{Rigol}},
  \bibinfo{journal}{Phys. Rev. A} \textbf{\bibinfo{volume}{82}},
  \bibinfo{pages}{043601} (\bibinfo{year}{2010}).

\bibitem[{\citenamefont{Mazets and Schmiedmayer}(2010)}]{Mazets2010}
\bibinfo{author}{\bibfnamefont{I.~E.} \bibnamefont{Mazets}} \bibnamefont{and}
  \bibinfo{author}{\bibfnamefont{J.}~\bibnamefont{Schmiedmayer}},
  \bibinfo{journal}{New J. Phys.} \textbf{\bibinfo{volume}{12}},
  \bibinfo{pages}{055023} (\bibinfo{year}{2010}).

\bibitem[{\citenamefont{Braaten and Hammer}(2007)}]{Braaten2007}
\bibinfo{author}{\bibfnamefont{E.}~\bibnamefont{Braaten}} \bibnamefont{and}
  \bibinfo{author}{\bibfnamefont{H.-W.} \bibnamefont{Hammer}},
  \bibinfo{journal}{Ann. Phys.} \textbf{\bibinfo{volume}{322}},
  \bibinfo{pages}{120} (\bibinfo{year}{2007}).

\bibitem[{\citenamefont{Busch et~al.}(1998)\citenamefont{Busch, Englert,
  Rz\c{a}\.zewski, and Wilkens}}]{Busch1998}
\bibinfo{author}{\bibfnamefont{T.}~\bibnamefont{Busch}},
  \bibinfo{author}{\bibfnamefont{B.}~\bibnamefont{Englert}},
  \bibinfo{author}{\bibfnamefont{K.}~\bibnamefont{Rz\c{a}\.zewski}},
  \bibnamefont{and} \bibinfo{author}{\bibfnamefont{M.}~\bibnamefont{Wilkens}},
  \bibinfo{journal}{Found. of Phys.} \textbf{\bibinfo{volume}{28}},
  \bibinfo{pages}{549} (\bibinfo{year}{1998}).

\end{thebibliography}

\end{document}